\documentclass[aps,prd,twocolumn,superscriptaddress,nofootinbib,
 amsmath,amssymb,floatfix,longbibliography]{revtex4-2}
\usepackage[T1]{fontenc}
\usepackage{lmodern}
\usepackage{graphicx}
\usepackage{amsmath,amssymb,mathrsfs,bm}
\usepackage{hyperref}
\usepackage{xcolor}
\hypersetup{colorlinks=true,citecolor=blue,linkcolor=blue,urlcolor=blue}

\begin{document}
\title{Polarization Signatures of Mixed Scalar–Pseudoscalar ALP–Photon Couplings in Intense Laser Pulses}
\author{Musajan Kadir}
\affiliation{Key Laboratory of Beam Technology of the Ministry of Education, and School of Physics and Astronomy, Beijing Normal University, Beijing 100875, China}
\author{Hong-Hao Fan}
\affiliation{School of Mathematics and Physics, Xinjiang Hetian College, Hetian 848000, China}
\author{Ying Wu}
\affiliation{Key Laboratory of Beam Technology of the Ministry of Education, and School of Physics and Astronomy, Beijing Normal University, Beijing 100875, China}
\author{Qi-Rong Wang}
\affiliation{Key Laboratory of Beam Technology of the Ministry of Education, and School of Physics and Astronomy, Beijing Normal University, Beijing 100875, China}
\author{Bai-Song Xie}
\email{bsxie@bnu.edu.cn}
\affiliation{Key Laboratory of Beam Technology of the Ministry of Education, and School of Physics and Astronomy, Beijing Normal University, Beijing 100875, China}
\affiliation{Institute of Radiation Technology, Beijing Academy of Science and Technology, Beijing 100875, China}
\date{\today}

\begin{abstract}
We investigate photon-to-axion-like-particle (ALP) conversion in a finite, linearly polarized
laser pulse for a single ALP field with simultaneous scalar-type and
pseudoscalar-type couplings to photons. Varying the collision angle
between the strong pulse and the probe provides a continuous
kinematic scan of the resonant ALP mass. For fixed kinematics in this
setting, the scalar-type and pseudoscalar-type ALP conversion amplitudes
interfere coherently, making both the photon-to-ALP conversion probability
and the polarization rotation depend on the difference between
the probe polarization angle $\alpha$ and the coupling angle
$\theta_g$. Conversion is maximal at
$|\alpha-\theta_g|=90^\circ$, where the polarization rotation
vanishes, while the rotation magnitude peaks at
$\alpha-\theta_g=45^\circ$ modulo $90^\circ$, where
conversion is half-maximal. These complementary polarization-angle
dependences provide polarization-resolved signatures of mixed
scalar--pseudoscalar photon couplings.
\end{abstract}

\maketitle

\section{Introduction}
\label{sec:introduction}

The Peccei--Quinn mechanism provides a dynamical solution to the
strong CP problem of quantum chromodynamics and predicts the axion
\cite{PecceiQuinn1977,Weinberg1978,Wilczek1978,
	DiCortona2016,DiLuzio2020}.
The axion is also a well-motivated dark-matter candidate, and its
coupling to photons underlies a broad range of laboratory,
astrophysical, and cosmological searches
\cite{Sikivie1983,RaffeltStodolsky1988,DuffyVanBibber2009,
	Marsh2016,Baryakhtar2025AxionReview}.
More generally, axion-like particles (ALPs) arise in many extensions
of the Standard Model without obeying the QCD-axion relation between
mass and coupling strength
\cite{JaeckelRingwald2010,IrastorzaRedondo2018,
	Choi2021,BauerNeubertThamm2017,BrivioEtAl2017}.
Astrophysical and cosmological observations constrain ALP--photon
couplings through stellar evolution, spectral measurements, and
polarization observables
\cite{Payez2015,Fordham:2024,MAGIC:2024,Zhang:2024,Namikawa:2025}.

Laboratory experiments search for axions and ALPs and test their
interactions with photons over a broad range of parameters
\cite{PDG2024Axions,CAST2017,ADMX2020,Kim:2024axion,Heinze2024,
	GavilanMartin:2025}. In external electromagnetic fields, real
photon-to-ALP conversion can induce polarization-dependent attenuation
(dichroism), whereas photon--ALP mixing can produce dispersive
modifications to light propagation and give rise to birefringence
\cite{RaffeltStodolsky1988,Liao2007,TamYang2012}.
These absorptive and dispersive effects can appear as changes in
intensity, polarization orientation, and ellipticity
\cite{Cameron1993,ChenMeiNi2007,Zavattini2008,DellaValle2016}.

High-intensity laser pulses provide a controlled laboratory setting
in which probe photons can exchange energy and momentum with a
background electromagnetic field, enabling conversion into on-shell
ALPs.
Early laser-based studies identified resonantly enhanced
photon-to-ALP conversion at masses set by the optical frequencies and
the associated possibility of frequency-shifted probe photons
\cite{DobrichGies2010}.
Subsequent work investigated scalar-type and pseudoscalar-type interactions of a spin-zero field as separate cases and derived their dichroic and
birefringent signatures, including real-particle production near
resonance
\cite{VillalbaDiPiazza2013}.
Further studies incorporated QED vacuum-polarization effects in the
pseudoscalar case
\cite{VillalbaChavez2014}
and generalized the polarimetric treatment to arbitrary plane-wave
pulses, quantifying the dependence of polarization rotation and
ellipticity on the pulse profile
\cite{VillalbaPodszusMuller2017}.

These studies establish the strong-laser framework for ALP conversion
and polarization effects, but have generally treated scalar-type and
pseudoscalar-type photon couplings as separate cases. In the present
work, both interactions couple to the same ALP field
and therefore contribute coherently to the same photon-to-ALP
conversion amplitude. We study this interference in a finite,
linearly polarized laser pulse, treating the probe polarization angle
and collision angle as complementary controls. This mixed-coupling setup introduces an interference-induced polarization structure that does not arise when the scalar and pseudoscalar interactions are treated separately.

Three features are central to the analysis. First, coherent
interference between the scalar-type and pseudoscalar-type amplitudes
selects orthogonal probe-polarization directions of vanishing and
maximal conversion, allowing polarization-angle scans to probe the
relative scalar--pseudoscalar coupling composition. Second, varying
the collision angle provides a continuous kinematic scan of the
resonantly enhanced ALP mass. Third, polarization-selective depletion
through real ALP production induces dichroic rotation: the rotation
vanishes along both the dark and bright polarization directions and
reaches its largest magnitude at intermediate orientations, where the
conversion probability is half of its polarization maximum. We focus
on this absorptive rotation and do not include dispersive
birefringence.

The paper is organized as follows.
Section~\ref{sec:model} introduces the effective interaction and the
probe--laser collision geometry.
Section~\ref{sec:conversion} derives the polarization structure of the
conversion probability and the finite-pulse resonance condition.
Section~\ref{sec:polarization_rotation} introduces the
coupling-selected polarization basis and analyzes the resulting
dichroic rotation.
Our conclusions are summarized in Sec.~\ref{sec:conclusions}.

\section{ALP--photon interaction and collision geometry}
\label{sec:model}

Mixed scalar--pseudoscalar photon couplings have been explored in
collider and photon-fusion studies, where angular and polarization
correlations provide sensitivity to the relative scalar and
pseudoscalar interaction components
\cite{Bolognesi2012,DingLiuSong2026,ChangZhao2026}.
By contrast, strong-laser analyses have generally treated scalar-type
and pseudoscalar-type ALP couplings as separate cases; for example,
Ref.~\cite{VillalbaDiPiazza2013} studied their respective optical
signatures in a counterpropagating probe--laser configuration. In the
present framework, both interaction structures couple to the same ALP
field and therefore contribute coherently to the same photon-to-ALP
conversion process. We retain the finite-pulse character of the strong
field while allowing the collision geometry to vary rather than
restricting the analysis to a fixed counterpropagating configuration.

\subsection{ALP--photon interaction}
\label{subsec:interaction}

We consider a single real ALP field $\phi(x)$ coupled to photons
through simultaneous scalar-type and pseudoscalar-type interactions,
with coupling strengths $g_s$ and $g_p$, respectively. It is
convenient to parameterize the two couplings as
\begin{equation}
	\begin{aligned}
		g_s&=g\sin\theta_g,\qquad
		g_p=g\cos\theta_g,\\
		g&=\sqrt{g_s^2+g_p^2},\qquad
		0^\circ\leq\theta_g\leq90^\circ,
	\end{aligned}
	\label{eq:couplings}
\end{equation}
where $g$ sets the overall coupling strength, while $\theta_g$
determines the relative scalar-type and pseudoscalar-type coupling
composition. The endpoints $\theta_g=0^\circ$ and $90^\circ$
correspond to the pure pseudoscalar-type and pure scalar-type limits,
respectively, whereas intermediate values describe the mixed-coupling
regime considered below.

It is noticed that in present work we consider only the absorptive effect associated
with real photon-to-ALP conversion. The QED vacuum birefringence and other dispersive modifications of the outgoing
probe polarization are not included in the present calculation. Accordingly the involved term of effective Lagrangian is associated to the ALP--photon interaction as below
\begin{equation}
	\mathscr L_{\rm int}
	=
	-\frac{1}{4\pi}g_s\phi\,\mathscr F
	-\frac{1}{4\pi}g_p\phi\,\mathscr G,
	\label{eq:lagrangian}
\end{equation}
where $\phi$ is treated as a generic ALP field without assigning a
definite CP parity. If $\phi$ is assigned the conventional
pseudoscalar CP parity, the $g_p$ interaction is CP conserving,
whereas the $g_s$ interaction is CP violating. The electromagnetic
invariants are defined by
\begin{align}
	\mathscr F
	&=\frac14F_{\mu\nu}F^{\mu\nu}
	=-\frac12\left(\mathbf E^2-\mathbf B^2\right),
	\nonumber\\
	\mathscr G
	&=\frac14F_{\mu\nu}\widetilde F^{\mu\nu}
	=-\mathbf E\cdot\mathbf B,
	\label{eq:invariants}
\end{align}
with the dual tensor
$\widetilde F^{\mu\nu}
=\frac12\epsilon^{\mu\nu\rho\sigma}F_{\rho\sigma}$
and $\epsilon^{0123}=+1$.
\subsection{Collision geometry and polarization basis}
\label{subsec:geometry}

\begin{figure}[htbp]
	\centering
	\includegraphics[width=0.75\columnwidth]{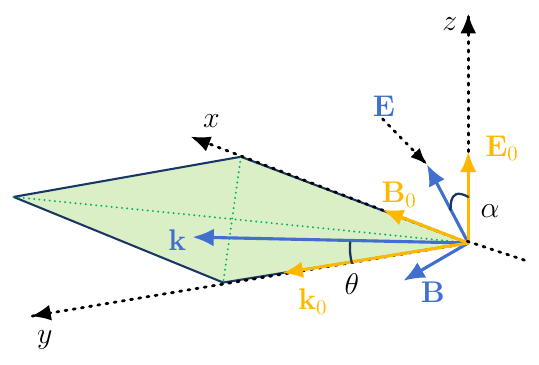}
	\caption{Collision geometry and polarization convention. The
		strong-laser quantities
		$(\mathbf{k}_0,\mathbf{E}_0,\mathbf{B}_0)$ are shown in
		gold, while the probe quantities
		$(\mathbf{k},\mathbf{E},\mathbf{B})$ are shown in blue.
		The strong laser propagates along $+y$, with its electric
		field polarized along $+z$, while the probe wave vector lies
		in the $x$--$y$ plane and forms the collision angle $\theta$
		with the laser. The angle $\alpha$ specifies the probe
		polarization relative to the strong-laser electric-field
		direction.}
	\label{fig:geometry}
\end{figure}

The setup consists of a strong, linearly polarized laser pulse
colliding with a weak probe photon at a variable collision angle. We denote by $\theta$ the angle between their propagation directions and take
$0^\circ<\theta\leq180^\circ$, with $\theta=180^\circ$
corresponding to a head-on collision. The strong laser propagates
along $+y$, with its electric field polarized along $+z$, while the
probe wave vector lies in the $x$--$y$ plane. The collision geometry
and polarization convention are illustrated in
Fig.~\ref{fig:geometry}. The corresponding wave vectors are
parameterized as
\begin{equation}
	k_0^\mu=\omega_0(1,0,1,0),\qquad
	k^\mu=\omega(1,\sin\theta,\cos\theta,0),
	\label{eq:momenta}
\end{equation}
where $\omega_0$ and $\omega$ are the angular frequencies of the
strong-laser carrier and the probe photon, respectively. Since
$k\!\cdot k_0=\omega\omega_0(1-\cos\theta)$, the co-propagating
limit $\theta=0^\circ$ gives a vanishing probe--background invariant
and is therefore excluded. Varying $\theta$ continuously changes the
collision invariant and hence the phase-matching condition for
photon-to-ALP conversion.

For the probe, we introduce an orthonormal polarization basis
transverse to $\mathbf{k}$,
\begin{equation}
	\hat{\mathbf e}_{\parallel}=\hat{\mathbf z},\qquad
	\hat{\mathbf e}_{\perp}
	=\cos\theta\,\hat{\mathbf x}
	-\sin\theta\,\hat{\mathbf y},
	\label{eq:polarization_basis}
\end{equation}
where $\hat{\mathbf e}_{\parallel}$ coincides with the strong-laser
electric-field polarization, while $\hat{\mathbf e}_{\perp}$ is the
orthogonal transverse direction. The probe polarization angle is
denoted by $\alpha$, with
$0^\circ\leq\alpha\leq180^\circ$.

The unit vector along the probe propagation direction is
\begin{equation}
	\hat{\mathbf n}
	=\frac{\mathbf k}{|\mathbf k|}
	=\sin\theta\,\hat{\mathbf x}
	+\cos\theta\,\hat{\mathbf y}.
	\label{eq:probe_direction}
\end{equation}
For a plane-wave probe,
$\mathbf b_\alpha
=\hat{\mathbf n}\times\boldsymbol{\epsilon}_\alpha$,
so the electric- and magnetic-field polarization unit vectors are
\begin{align}
	\boldsymbol{\epsilon}_\alpha
	&=\cos\alpha\,\hat{\mathbf z}
	+\sin\alpha\left(
	\cos\theta\,\hat{\mathbf x}
	-\sin\theta\,\hat{\mathbf y}\right),
	\nonumber\\
	\mathbf b_\alpha
	&=\cos\alpha\left(
	\cos\theta\,\hat{\mathbf x}
	-\sin\theta\,\hat{\mathbf y}\right)
	-\sin\alpha\,\hat{\mathbf z}.
	\label{eq:probe_polarization}
\end{align}
By construction,
$(\boldsymbol{\epsilon}_\alpha,\mathbf b_\alpha,\hat{\mathbf n})$
forms a right-handed orthonormal triad.

Within this geometry, the angles $\theta$ and $\alpha$ control distinct aspects of the
conversion process. The collision angle changes the
probe--background invariant and therefore the kinematic phase
matching, whereas the probe polarization angle changes the
polarization projections entering the scalar-type and
pseudoscalar-type contributions to the conversion amplitude. This
separation provides independent kinematic and polarization controls
for the analysis below.

\section{Photon-to-ALP conversion and finite-pulse resonance}
\label{sec:conversion}

\subsection{Coherent conversion amplitude and probability}
\label{subsec:conversion_amplitude}

\begin{figure}[htbp]
	\centering
	\includegraphics[width=0.30\textwidth]{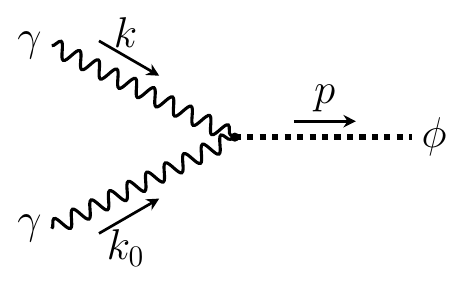}
	\caption{Leading-order photon-to-ALP conversion in the prescribed
		classical laser background. A quantized probe photon $\gamma(k)$
		converts into an on-shell ALP $\phi(p)$ through its interaction with
		the external field.}
	\label{fig:conversion}
\end{figure}

We consider the leading-order conversion of a quantized probe photon
into a real, on-shell ALP in the prescribed classical laser background
shown in Fig.~\ref{fig:conversion}. The strong pulse is treated as an
external field, with depletion and backreaction neglected, while the
probe photon and produced ALP are quantized. For ease of comparison with the finite-pulse treatment of
Ref.~\cite{VillalbaDiPiazza2013}, we follow the same pulse-phase and
Fourier-transform conventions where applicable. We model the background as
a linearly polarized pulsed plane wave,
\begin{equation}
 \mathbf E_0(\varphi)=\mathcal E_0(\varphi)\,\hat{\mathbf z},\qquad
 \mathbf B_0(\varphi)=\mathcal E_0(\varphi)\,\hat{\mathbf x},
 \label{eq:background_field}
\end{equation}
where
$\mathcal E_0(\varphi)=E_0 f(\varphi)
\cos(\varphi+\varphi_{\rm CEP})$ and
$\varphi=k_0\!\cdot x=\omega_0(t-y)$. Here $E_0$ is the peak field
amplitude, $f(\varphi)$ is the pulse envelope, and
$\varphi_{\rm CEP}$ is the carrier-envelope phase. This pulsed plane-wave setup is a standard idealization in strong-field
ALP production and conversion studies
\cite{DobrichGies2010,VillalbaDiPiazza2013,DillonKing2018}.

Writing $F_{\mu\nu}=F_{0,\mu\nu}+f_{\mu\nu}$ and keeping only terms
linear in the quantized probe field gives
\begin{equation}
 \mathscr L_{\rm int}^{(1)}
 =-\frac{\phi}{8\pi}
 \left(g_sF_{0,\mu\nu}f^{\mu\nu}
 +g_p\widetilde F_{0,\mu\nu}f^{\mu\nu}\right),
 \label{eq:mixing}
\end{equation}
where $F_{0,\mu\nu}$ is the background-field tensor and
$f_{\mu\nu}$ is the quantized probe-field tensor. For the
geometry of Sec.~\ref{subsec:geometry}, the scalar-type and
pseudoscalar-type contributions enter the same transition amplitude.
We define the polarization-dependent coherent coupling factor as
\begin{equation}
 \Gamma_\alpha
 \equiv g_s\cos\alpha-g_p\sin\alpha
 =-g\sin(\alpha-\theta_g),
 \label{eq:amplitude_factor}
\end{equation}
thus $\Gamma_\alpha$ has the same dimensions as $g_s$ and $g_p$ and
multiplies the common kinematic and pulse factor in the conversion
amplitude. Unlike the pure scalar-type or pure pseudoscalar-type cases, both
couplings contribute to the same transition amplitude and therefore
interfere coherently.

The finite-pulse dependence is described by the Fourier transform
$G$ of the envelope and by the carrier-resolved spectral amplitude
$H$,
\begin{align}
 G(\nu)
 &=\int_{-\infty}^{\infty}d\varphi\,
 f(\varphi)e^{i\nu\varphi},
 \nonumber\\
 H(\rho)
 &=e^{i\varphi_{\rm CEP}}G(\rho+1)
 +e^{-i\varphi_{\rm CEP}}G(\rho-1),
 \label{eq:pulse_spectrum}
\end{align}
where \(\nu\) is the dimensionless Fourier variable conjugate to the
laser phase \(\varphi\). The function \(H\) is introduced as a compact
notation for the two carrier-shifted Fourier components. The
dimensionless on-shell light-front momentum-transfer parameter is
\begin{equation}
 \rho
 =\frac{m_\phi^2}{2k\!\cdot k_0}
 =\frac{m_\phi^2}
 {4\omega\omega_0\sin^2(\theta/2)}.
 \label{eq:rho}
\end{equation}
The condition $\rho=1$ corresponds to matching the
carrier-frequency component of the pulse. The intermediate $S$-matrix, light-front, and
phase-space steps are derived explicitly in
Appendix~\ref{app:conversion_derivation}. The resulting conversion
probability is
\begin{align}
 P_{\gamma\to\phi}
 &=\frac{I_{\rm amp}}{16\omega_0^2}
 \sin^2\frac{\theta}{2}\,
 |g_s\cos\alpha-g_p\sin\alpha|^2
 |H(\rho)|^2
 \nonumber\\
 &=P_{\max}(\theta,m_\phi)
 \sin^2(\alpha-\theta_g),
 \label{eq:probability}
\end{align}
where $P_{\max}(\theta,m_\phi)$ is the maximum over the probe
polarization angle, attained when $|\sin(\alpha-\theta_g)|=1$:
\begin{equation}
 P_{\max}(\theta,m_\phi)
 =
 \frac{g^2I_{\rm amp}}{16\omega_0^2}
 \sin^2\frac{\theta}{2}|H(\rho)|^2.
 \label{eq:max_probability}
\end{equation}
Here $I_{\rm amp}=E_0^2/(4\pi)$ denotes the peak intensity parameter
defined from the field amplitude and corresponds to $I_0$ in
Ref.~\cite{VillalbaDiPiazza2013}. For a many-cycle linearly polarized pulse,
$I_{\rm amp}=2I_{\rm av}$, where $I_{\rm av}$ is the cycle-averaged
peak intensity. In the head-on pure-coupling limit,
Eq.~\eqref{eq:probability} reduces to the finite-pulse conversion
probability of Ref.~\cite{VillalbaDiPiazza2013}; the detailed
normalization correspondence is shown in
Appendix~\ref{app:conversion_derivation}.

Equation~\eqref{eq:probability} makes the physical structure explicit:
the pulse spectrum and collision geometry determine the common
kinematic response, while the coherent scalar--pseudoscalar
combination determines the polarization dependence through
$\alpha-\theta_g$. For fixed kinematics, pulse parameters, and
coupling composition, the conversion vanishes at $\alpha=\theta_g$
and is maximal for the orthogonal direction
$\alpha=\theta_g+90^\circ$, where
$P_{\gamma\to\phi}=P_{\max}$. These two orthogonal polarization
directions provide a natural basis for the polarization analysis
developed in Sec.~\ref{sec:polarization_rotation}.

\subsection{Finite-pulse resonance}
\label{subsec:finite_pulse}

For a many-cycle pulse, the Fourier transform $G(\nu)$ of the
envelope is sharply localized near $\nu=0$. Since $\rho\geq0$,
the term $G(\rho-1)$ is resonantly enhanced near $\rho=1$, whereas
$G(\rho+1)$ remains off resonance.
The corresponding carrier-frequency phase-matching condition is
therefore
\begin{equation}
 m_{\rm res}(\theta)
 =2\sqrt{\omega\omega_0}\sin\frac{\theta}{2}.
 \label{eq:resonance}
\end{equation}
Physically, varying the collision angle changes
the light-front momentum available from the laser and therefore moves
the resonant ALP mass. A head-on collision gives the largest resonant
mass, whereas smaller $\theta$ shifts the resonance downward and also
reduces the factor $\sin^2(\theta/2)$ in $P_{\max}$.

\begin{table}[htbp]
	\caption{Benchmark parameters used for
		Figs.~\ref{fig:polarization_maps}--\ref{fig:polarization_rotation}.}
	\label{tab:parameters}
	\begin{ruledtabular}
		\begin{tabular}{@{}lcc@{}}
			Parameter & Symbol & Value \\
			\hline
			Strong-laser carrier frequency & $\omega_0$ & $1.55~{\rm eV}$ \\
			Probe-photon frequency & $\omega$ & $3.10~{\rm eV}$ \\
			Gaussian width parameter & $N$ & $5.6$ \\
			Intensity-envelope FWHM & $N_{I,{\rm FWHM}}$ & $\simeq9.32$ cycles \\
			Carrier-envelope phase & $\varphi_{\rm CEP}$ & $0$ \\
			Peak intensity parameter & $I_{\rm amp}$ &
			$10^{25}~{\rm W/cm^2}$ \\
			Coupling magnitude & $g$ & $1.0\times10^{-7}~{\rm GeV}^{-1}$ \\
		\end{tabular}
	\end{ruledtabular}
\end{table}

For the numerical calculations, we take the pulse envelope to be
\begin{equation}
 f(\varphi)=\exp\!\left[-\frac{\varphi^2}{2\Delta\varphi^2}\right],
 \qquad \Delta\varphi=2\pi N.
 \label{eq:envelope}
\end{equation}
Here $\Delta\varphi$ is the Gaussian envelope width in laser phase and
$N=\Delta\varphi/(2\pi)$ is the corresponding dimensionless width
parameter measured in carrier cycles. Its Fourier transform is
$G(\nu)=\sqrt{2\pi}\Delta\varphi
\exp[-\Delta\varphi^2\nu^2/2]$. The intensity full width at half
maximum, expressed in optical cycles, is denoted
$N_{I,{\rm FWHM}}$ in Table~\ref{tab:parameters}.

For the numerical illustrations, we use the benchmark parameters
summarized in Table~\ref{tab:parameters}. The ultrashort, high-power
laser regime is motivated by facilities such as ELI-NP, while the
adopted focal intensity is representative of the higher intensities
considered for XCELS-class systems \cite{Ur2025,Khazanov2023}.
These parameters are chosen to illustrate the polarization and
resonance structure rather than to model a specific experimental
configuration. Together with Eq.~\eqref{eq:resonance}, the benchmark
frequencies determine the resonant mass scale explored below.

For comparison, the 2024 CAST vacuum result constrains the
conventional pseudoscalar axion--photon coupling to
$g_{a\gamma}<5.8\times10^{-11}\,{\rm GeV}^{-1}$ at 95\% C.L. for
$m_a\lesssim0.02~{\rm eV}$, while the ${}^3\mathrm{He}$ buffer-gas program
covered $0.64$--$1.17~{\rm eV}$ with an average upper limit of
$3.3\times10^{-10}\,{\rm GeV}^{-1}$ at 95\% C.L.
\cite{CAST2024,CAST2014BufferGas}. These limits assume a pure
pseudoscalar axion--photon interaction and therefore do not directly
constrain the mixed scalar--pseudoscalar coupling magnitude $g$
used here. Moreover, the CAST mass coverage extends only to about
$1.17~{\rm eV}$ and therefore does not cover the full several-eV
mass interval considered here. We use
$g=10^{-7}\,{\rm GeV}^{-1}$ as a reference coupling for the
numerical illustrations. This choice fixes the overall signal scale
while leaving the polarization and resonance structure unchanged; it
is not intended as a phenomenological benchmark.

\begin{figure}[htbp]
	\centering
	\includegraphics[width=0.95\columnwidth]{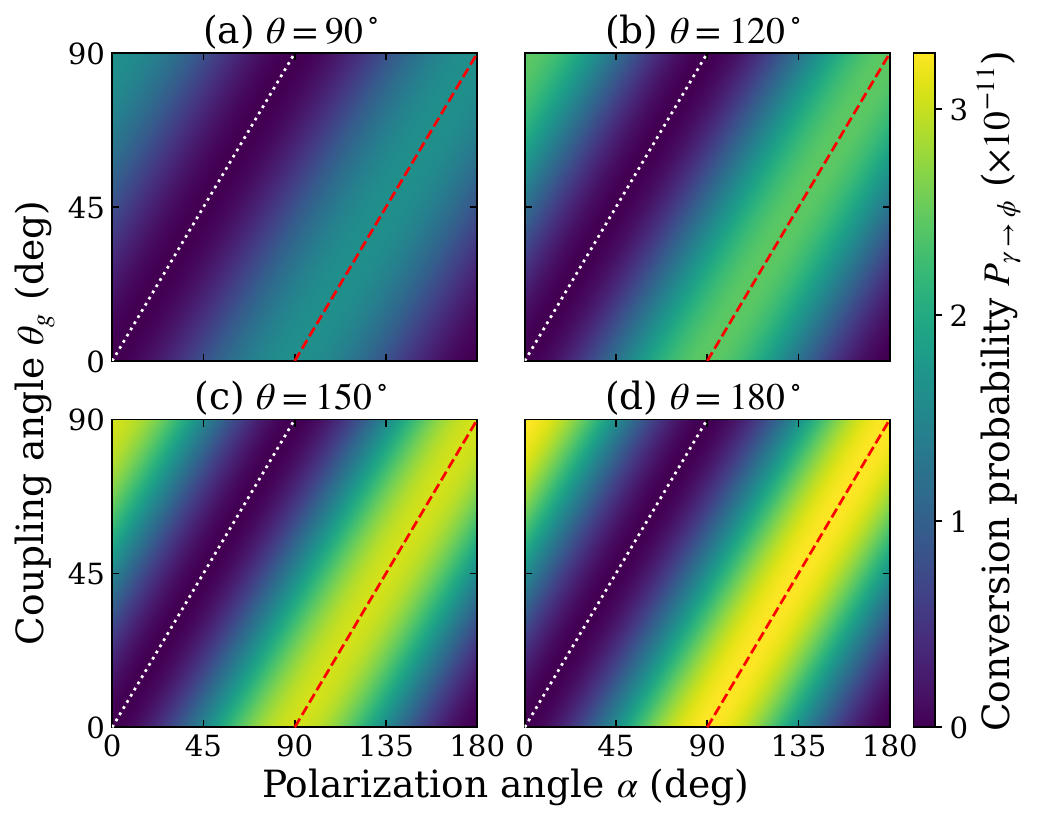}
	\caption{Photon-to-ALP conversion probability
		$P_{\gamma\to\phi}$ as a function of the probe polarization
		angle $\alpha$ and coupling angle $\theta_g$ for
		$\theta=90^\circ$, $120^\circ$, $150^\circ$, and $180^\circ$.
		In each panel, $m_\phi=m_{\rm res}(\theta)$, corresponding to
		$\rho=1$. The white dotted and red dashed lines indicate the
		zero-conversion branch $\alpha=\theta_g$ and the
		maximal-conversion branch $\alpha=\theta_g+90^\circ$,
		respectively. The remaining parameters are
		given in Table~\ref{tab:parameters}.}
	\label{fig:polarization_maps}
\end{figure}

Equations~\eqref{eq:probability} and \eqref{eq:resonance}, together
with the benchmark parameters in Table~\ref{tab:parameters}, provide
the basis for the numerical results below. Rather than varying all
parameters simultaneously, we exploit the factorized form of the
conversion probability to separate the polarization response from the
mass--angle resonance structure. We first impose the carrier-frequency
resonance, $m_\phi=m_{\rm res}(\theta)$, and scan the polarization
variables; we then fix the polarization on the maximal-conversion
branch and scan the $(m_\phi,\theta)$ plane to expose the remaining
kinematic and finite-pulse spectral dependence.

Figure~\ref{fig:polarization_maps} shows the first of these two
representations. In each panel we set
$m_\phi=m_{\rm res}(\theta)$, so that $\rho=1$ and the spectral factor
$|H(1)|^2$ is common to all collision angles. This removes spectral
detuning, leaving the coupling-dependent polarization pattern together
with the geometric $\sin^2(\theta/2)$ scaling.

The zero-conversion branch follows from the cancellation of the coherent
coupling factor $\Gamma_\alpha$; for intermediate $\theta_g$, this
cancellation corresponds to destructive interference between the
scalar-type and pseudoscalar-type conversion amplitudes. The orthogonal
branch gives the maximum conversion probability.
Thus $\theta_g$ fixes the orientation of the polarization pattern,
while $\theta$ changes its overall magnitude through $\sin^2(\theta/2)$ and simultaneously selects the resonant ALP
mass.

\begin{figure}[htbp]
	\centering
	\includegraphics[width=0.95\columnwidth]{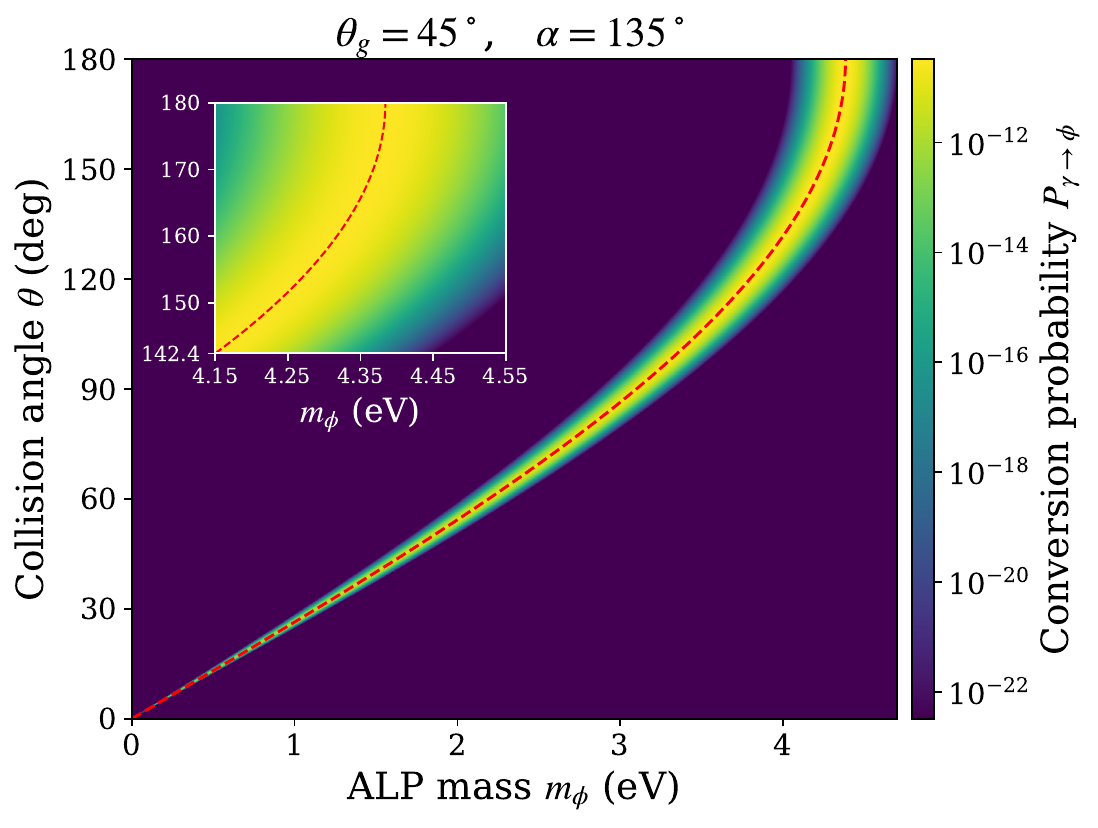}
	\caption{Photon-to-ALP conversion probability
		$P_{\gamma\to\phi}$ in the $(m_\phi,\theta)$ plane for the
		maximal-conversion choice
		$\theta_g=45^\circ$ and $\alpha=135^\circ$.
		The red dashed curve denotes the carrier-frequency
		phase-matching condition $\rho=1$. The inset enlarges the
		head-on endpoint region. The remaining parameters are listed
		in Table~\ref{tab:parameters}.}
	\label{fig:mass_angle_resonance}
\end{figure}

To expose the complementary mass--angle response, we fix
$\theta_g=45^\circ$ and $\alpha=135^\circ$, which lies on the
maximal-conversion branch and gives
$\sin^2(\alpha-\theta_g)=1$. With the polarization factor fixed at
unity, Fig.~\ref{fig:mass_angle_resonance} displays the remaining
kinematic and spectral dependence.

The enhanced ridge in Fig.~\ref{fig:mass_angle_resonance} follows the
carrier-frequency phase-matching condition $\rho=1$ and hence the
resonant mass given by Eq.~\eqref{eq:resonance}. It is therefore the
continuous $(m_\phi,\theta)$ locus of the resonant
maximal-conversion configurations sampled in
Fig.~\ref{fig:polarization_maps}. At $\theta=90^\circ$, $120^\circ$,
$150^\circ$, and $180^\circ$, the points on this ridge reproduce the
maximal-conversion values of the corresponding panels in
Fig.~\ref{fig:polarization_maps}, demonstrating the consistency
between the polarization and mass--angle representations of
Eq.~\eqref{eq:probability}.

Along the resonant ridge, decreasing $\theta$ shifts the resonance to
smaller masses and reduces the peak conversion probability through the
geometric factor $\sin^2(\theta/2)$, while the spectral factor remains
fixed at $|H(1)|^2$. The largest carrier-frequency resonant mass occurs
for a head-on collision. For the benchmark frequencies in Table~\ref{tab:parameters}, the
corresponding head-on endpoint is
$m_{\rm res}(180^\circ)=4.384~{\rm eV}$, above which no
carrier-frequency $\rho=1$ solution exists.

The inset of Fig.~\ref{fig:mass_angle_resonance} resolves this
head-on endpoint. Although the $\rho=1$ curve terminates there, the
finite pulse bandwidth produces a finite-width high-mass tail beyond
the carrier-frequency endpoint. This tail is the off-resonant
high-mass wing of the same resonance rather than a second resonance
branch.

\section{Dichroic rotation of the probe polarization}
\label{sec:polarization_rotation}

Equation~\eqref{eq:amplitude_factor} shows that, at leading order,
the mixed coupling selects a single linear combination of the two
transverse probe polarizations. The two orthogonal directions
identified in Sec.~\ref{subsec:conversion_amplitude} therefore define
a natural polarization basis: we refer to the maximal-conversion
direction as the bright mode and to the zero-conversion direction as
the dark mode. Their orthonormal unit vectors are
\begin{align}
	\hat{\mathbf e}_b
	&=\sin\theta_g\,\hat{\mathbf e}_{\parallel}
	-\cos\theta_g\,\hat{\mathbf e}_{\perp},
	\nonumber\\
	\hat{\mathbf e}_d
	&=\cos\theta_g\,\hat{\mathbf e}_{\parallel}
	+\sin\theta_g\,\hat{\mathbf e}_{\perp}.
	\label{eq:bright_dark_modes}
\end{align}
The incident probe polarization can then be decomposed as
\begin{equation}
	\boldsymbol\epsilon_\alpha
	=-\sin(\alpha-\theta_g)\hat{\mathbf e}_b
	+\cos(\alpha-\theta_g)\hat{\mathbf e}_d.
	\label{eq:bright_dark_decomposition}
\end{equation}
Since $\Gamma_\alpha=g\,\hat{\mathbf e}_b\!\cdot\boldsymbol\epsilon_\alpha=-g\sin(\alpha-\theta_g)$,
the conversion amplitude is determined entirely by the bright-mode
component, whereas the dark mode does not contribute to real,
on-shell ALP production at this order. The bright--dark basis thus
provides a direct geometric interpretation of the factor
$\sin^2(\alpha-\theta_g)$ in Eq.~\eqref{eq:probability}. For later
use, we denote the conversion probability of a purely bright input by
\begin{equation}
	P_b(\theta,m_\phi)\equiv P_{\max}(\theta,m_\phi),
\end{equation}
with $P_{\max}$ given in Eq.~\eqref{eq:max_probability}.

This bright--dark decomposition also makes the origin of the
conversion-induced polarization rotation transparent. Real ALP
production selectively depletes the bright component of the probe,
while the dark component remains unaffected at this order. This
polarization-dependent attenuation constitutes dichroism. When both
components are present, the resulting change in their relative
amplitudes rotates the polarization direction of the transmitted
probe.

The underlying mechanism is the same general dichroic effect familiar
from conventional photon--ALP mixing in external magnetic fields
\cite{Maiani1986,RaffeltStodolsky1988}. In the present case, however,
the bright and dark directions are determined by the coherent
scalar--pseudoscalar coupling composition, while the attenuation of
the bright mode is governed by the finite-pulse conversion probability
derived in Sec.~\ref{sec:conversion}.

\begin{figure}[htbp]
	\centering
	\includegraphics[width=0.95\columnwidth]{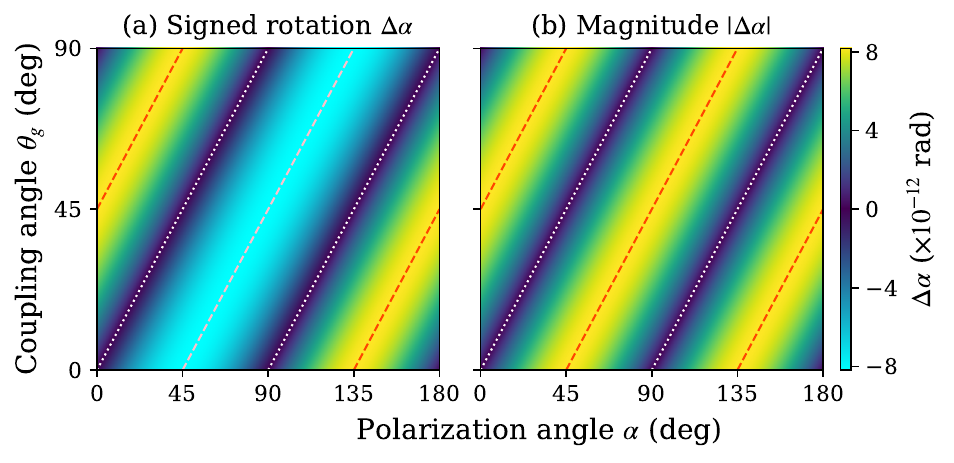}
	\caption{Dichroic rotation for a head-on collision at
		$m_\phi=m_{\rm res}(180^\circ)$. (a) Signed rotation
		$\Delta\alpha$. The dashed branches at
		$\alpha-\theta_g=45^\circ$ modulo $90^\circ$ mark extrema of
		the rotation magnitude; adjacent branches alternate between
		negative and positive rotation. The dotted branches at
		$\alpha-\theta_g=0^\circ$ modulo $90^\circ$ indicate zero
		rotation. (b) Rotation magnitude $|\Delta\alpha|$, for which
		the dashed branches mark the maxima and the dotted branches
		mark the zeros. The remaining parameters are listed in
		Table~\ref{tab:parameters}.}
	\label{fig:polarization_rotation}
\end{figure}

In the weak-conversion regime $P_b\ll1$, the bright-field amplitude is
reduced by $\sqrt{1-P_b}\simeq1-P_b/2$, while the dark component is
unchanged. Writing the change of probe polarization angle as
$\Delta\alpha\equiv\alpha_{\rm out}-\alpha$, one obtains
\begin{equation}
	\Delta\alpha
	=-\frac{P_b}{4}\sin[2(\alpha-\theta_g)].
	\label{eq:rotation}
\end{equation}
A short derivation is given in
Appendix~\ref{app:rotation_derivation}. Holding all other parameters fixed, so that $P_b$ is independent of $\alpha$, the same
result can be written as
\begin{equation}
	\Delta\alpha
	=
	-\frac14
	\frac{\partial P_{\gamma\to\phi}}{\partial\alpha},
	\label{eq:rotation_probability_relation}
\end{equation}
where $\alpha$ is understood in radians. The rotation magnitude is
maximal for
\begin{equation}
	\alpha-\theta_g
	=45^\circ \pmod {90^\circ},
	\qquad
	|\Delta\alpha|_{\rm max}
	=\frac{P_b}{4}.
	\label{eq:maximum_rotation}
\end{equation}
For the head-on resonant benchmark of
Table~\ref{tab:parameters},
$|\Delta\alpha|_{\rm max}\simeq8.09\times10^{-12}\,{\rm rad}$.
The corresponding angular dependence is shown in
Fig.~\ref{fig:polarization_rotation}.

Figs.~\ref{fig:polarization_maps} and
~\ref{fig:polarization_rotation} show complementary manifestations of
the same coupling-defined bright--dark structure. The dark direction
$\alpha=\theta_g$ and the orthogonal bright direction give,
respectively, zero and maximal conversion in
Fig.~\ref{fig:polarization_maps}, yet both are zero-rotation directions
in Fig.~\ref{fig:polarization_rotation}: a pure dark input is
unchanged, whereas a pure bright input is attenuated without changing
its polarization direction. Rotation requires both components and is
therefore largest when their incident amplitudes have equal magnitude,
$\alpha-\theta_g=45^\circ$ modulo $90^\circ$. In panel (a), the sign
of $\Delta\alpha$ distinguishes the two senses of rotation, while
panel (b) shows only the corresponding magnitude.

\section{Conclusions}
\label{sec:conclusions}

We have studied leading-order photon-to-ALP conversion in a finite,
linearly polarized laser pulse with simultaneous scalar-type and
pseudoscalar-type photon couplings and a variable collision geometry.

Three main results emerge. First, the two interaction contributions
enter the same transition amplitude and interfere coherently, leading
to the factorized polarization dependence
$P_{\gamma\to\phi}=P_{\max}\sin^2(\alpha-\theta_g)$.
The coupling composition therefore determines two orthogonal probe
polarizations corresponding to vanishing and maximal conversion,
respectively. Second, the finite-pulse resonance is controlled by the collision
geometry through
$m_{\rm res}(\theta)=2\sqrt{\omega\omega_0}\sin(\theta/2)$.
Varying the collision angle continuously shifts the resonantly
enhanced ALP mass while leaving the coupling-selected polarization
directions unchanged. The polarization maps and the mass--angle
resonance structure therefore provide complementary representations
of the same factorized conversion probability. Third, real on-shell ALP production induces dichroic polarization
rotation through polarization-dependent depletion of the transmitted
probe. In the weak-conversion regime,
$\Delta\alpha=-(P_b/4)\sin[2(\alpha-\theta_g)]$.
The rotation vanishes along both the zero-conversion and
maximal-conversion directions and reaches its maximum magnitude when
the two corresponding polarization components have equal incident
amplitudes. At these angles, the conversion probability is half of
its maximum. Conversion and
polarization rotation thus provide complementary polarization-resolved
signatures of the same coherent scalar--pseudoscalar interaction
structure.

The present analysis isolates the absorptive effect associated with
real ALP production in a pulsed plane-wave background with a
monochromatic probe. The framework can be extended to include
ALP-induced dispersive phases, QED vacuum birefringence, and more general strong-field polarization states.

\begin{acknowledgments}
    We thank Orkash Amat and Mamat Ali Bake for helpful discussions. This work is supported by the National Natural Science Foundation of China (NSFC) under Grants No. 12375240 and No. 12535015.
\end{acknowledgments}

\appendix

\section{Derivation of the photon-to-ALP conversion probability}
\label{app:conversion_derivation}

This appendix derives the conversion probability in
Eq.~\eqref{eq:probability} using the normalization conventions of the
main text. The scalar and pseudoscalar contributions are combined at
the amplitude level, the finite-pulse spacetime integral supplies the
spectral factor, and the ALP on-shell condition together with the
final-state phase-space integration determines the kinematic and
normalization factors.

At first order in the quantized probe field, the conversion amplitude
is determined by the linearized invariants
$\mathscr F^{(1)}=\mathbf B_0\!\cdot\mathbf B_\gamma
-\mathbf E_0\!\cdot\mathbf E_\gamma$ and
$\mathscr G^{(1)}=-(\mathbf E_0\!\cdot\mathbf B_\gamma
+\mathbf B_0\!\cdot\mathbf E_\gamma)$.
We use continuum-normalized one-particle states,
$\langle\mathbf p|\mathbf p'\rangle=(2\pi)^3
\delta^{(3)}(\mathbf p-\mathbf p')$, with the same normalization
convention for the incident photon. The required field matrix
elements are
\begin{align}
	\langle0|\mathbf E_\gamma(x)|\gamma(k,\alpha)\rangle
	&=i\sqrt{2\pi\omega}\,
	\boldsymbol\epsilon_\alpha e^{-ik\cdot x},
	\nonumber\\
	\langle0|\mathbf B_\gamma(x)|\gamma(k,\alpha)\rangle
	&=i\sqrt{2\pi\omega}\,
	\mathbf b_\alpha e^{-ik\cdot x},
	\nonumber\\
	\langle\phi(p)|\phi(x)|0\rangle
	&=\frac{e^{ip\cdot x}}{\sqrt{2\varepsilon_p}},
	\qquad
	\varepsilon_p=\sqrt{m_\phi^2+\mathbf p^2}.
	\label{eq:appA_field_matrix_elements}
\end{align}

Contracting the probe fields with the background geometry of
Eqs.~\eqref{eq:background_field} and \eqref{eq:probe_polarization}
gives
\begin{align}
	\langle\mathscr F^{(1)}\rangle
	&=-i\sqrt{2\pi\omega}\,
	\mathcal E_0(\varphi)(1-\cos\theta)
	\cos\alpha\,e^{-ik\cdot x},
	\nonumber\\
	\langle\mathscr G^{(1)}\rangle
	&=+i\sqrt{2\pi\omega}\,
	\mathcal E_0(\varphi)(1-\cos\theta)
	\sin\alpha\,e^{-ik\cdot x},
	\label{eq:appA_invariants}
\end{align}
where
$\langle\mathscr F^{(1)}\rangle\equiv
\langle0|\mathscr F^{(1)}(x)|\gamma(k,\alpha)\rangle$,
and analogously for $\mathscr G^{(1)}$.
The relative sign causes the scalar and pseudoscalar contributions
to combine coherently through
$\Gamma_\alpha=g_s\cos\alpha-g_p\sin\alpha$,
defined in Eq.~\eqref{eq:amplitude_factor}. The leading-order
$S$-matrix element is therefore, up to an overall phase,
\begin{align}
	S_{fi}^{(1)}
	&=-\frac{E_0}{4\pi}
	\sqrt{\frac{\pi\omega}{\varepsilon_p}}
	(1-\cos\theta)\Gamma_\alpha\,\mathcal I(q),
	\nonumber\\
	\mathcal I(q)
	&=\int d^4x\,
	f(\varphi)\cos(\varphi+\varphi_{\rm CEP})e^{iq\cdot x}.
	\label{eq:appA_smatrix}
\end{align}
Thus $\Gamma_\alpha$ contains the complete polarization dependence,
whereas the pulse profile and kinematics enter through
$\mathcal I(q)$. Here $q=p-k$ is the four-momentum transfer associated
with the photon-to-ALP transition, as illustrated in
Fig.~\ref{fig:conversion}.

To evaluate the pulse integral, we introduce the light-front
variables $x^\pm=t\pm y$ and $q^\pm=q^0\pm q^y$. Since
$\varphi=\omega_0x^-$, the integrations over $x$, $z$, and $x^+$
enforce $q_x=q_z=q^-=0$, while the remaining $x^-$ integration gives
the pulse spectrum defined in Eq.~\eqref{eq:pulse_spectrum}:
\begin{equation}
	\mathcal I(q)
	=(2\pi)^3\delta(q_x)\delta(q_z)\delta(q^-)
	\frac{H(\rho)}{2\omega_0},
	\qquad
	\rho=\frac{q^+}{2\omega_0}.
	\label{eq:appA_pulse_integral}
\end{equation}
The three conservation conditions imply
$p^\mu=k^\mu+\rho k_0^\mu$. Imposing $p^2=m_\phi^2$ then gives
\begin{equation}
	m_\phi^2=2\rho\,k\!\cdot k_0,
	\qquad
	\rho=
	\frac{m_\phi^2}{4\omega\omega_0\sin^2(\theta/2)},
	\label{eq:appA_rho}
\end{equation}
which is Eq.~\eqref{eq:rho}. Hence $\rho$ fixes the kinematic
momentum transfer, while $H(\rho)$ supplies the corresponding
finite-pulse spectral weight.

With continuum-normalized states, the conversion probability follows
from the squared transition amplitude, integrated over the final ALP
phase space and divided by the norm of the incident plane-wave state.
To make the normalization of the squared conservation distributions
explicit, we temporarily regulate the $x$, $z$, and $x^+$
integrations over finite intervals of lengths $L_x$, $L_z$, and
$L_+$. Defining
\begin{equation}
	\mathcal D(q)
	=(2\pi)^3\delta(q_x)\delta(q_z)\delta(q^-),
\end{equation}
the squared delta functions are understood as
\begin{align}
	[(2\pi)\delta(q_x)]^2
	&\rightarrow L_x(2\pi)\delta(q_x),
	\nonumber\\
	[(2\pi)\delta(q_z)]^2
	&\rightarrow L_z(2\pi)\delta(q_z),
	\nonumber\\
	[(2\pi)\delta(q^-)]^2
	&\rightarrow \frac{L_+}{2}(2\pi)\delta(q^-).
\end{align}
The factor $1/2$ in the last relation arises because $q^-$ is
Fourier conjugate to $x^+/2$. At fixed time, the normalization volume
of the incident state is $V=L_xL_+L_z$, so the regulator-dependent
volume factors cancel against the norm of the incoming plane-wave
state.

The transverse delta functions fix $p_x=k_x$ and $p_z=k_z$. For the
remaining longitudinal integration,
$p^-=\varepsilon_p-p_y$ and
$k^-=\omega(1-\cos\theta)$, so the constraint $p^-=k^-$ selects a
single root $p_y^\star$. Since
$\left|d(p^--k^-)/dp_y\right|_{p_y^\star}=k^-/\varepsilon_p$,
\begin{align}
	\delta(p^--k^-)
	&=
	\frac{\varepsilon_p}{k^-}
	\delta(p_y-p_y^\star),
	\nonumber\\
	\frac{1}{V}
	\int\frac{d^3p}{(2\pi)^3}|\mathcal D(q)|^2
	&\longrightarrow
	\frac{\varepsilon_p}{2k^-}.
	\label{eq:appA_phase_space}
\end{align}
Thus all regulator-dependent volume factors cancel, leaving the
finite longitudinal Jacobian together with the factor $1/2$ associated
with the light-front normalization.

Combining the $S$-matrix element in Eq.~\eqref{eq:appA_smatrix} with
the pulse integral in Eq.~\eqref{eq:appA_pulse_integral} and carrying
out the final-state phase-space integration using
Eq.~\eqref{eq:appA_phase_space}, with
$k^-=\omega(1-\cos\theta)$, gives
\begin{equation}
	P_{\gamma\to\phi}
	=\frac{I_{\rm amp}}{16\omega_0^2}
	\sin^2\frac{\theta}{2}\,
	|\Gamma_\alpha|^2|H(\rho)|^2,
	\qquad
	I_{\rm amp}=\frac{E_0^2}{4\pi}.
	\label{eq:appA_probability}
\end{equation}
For real couplings,
\begin{align}
	|\Gamma_\alpha|^2
	&=g_s^2\cos^2\alpha+g_p^2\sin^2\alpha
	-g_sg_p\sin(2\alpha)
	\nonumber\\
	&=g^2\sin^2(\alpha-\theta_g).
	\label{eq:appA_interference}
\end{align}
The first form displays the scalar--pseudoscalar interference
explicitly, while the second follows from Eq.~\eqref{eq:couplings}.
Equations~\eqref{eq:appA_probability} and
\eqref{eq:appA_interference} therefore yield the factorized
probability given in Eq.~\eqref{eq:probability}, with
$P_{\max}$ defined in Eq.~\eqref{eq:max_probability}.

For a head-on collision and a pure scalar or pseudoscalar coupling,
Eq.~\eqref{eq:appA_probability} agrees with the corresponding
linearly polarized finite-pulse conversion probability in
Ref.~\cite{VillalbaDiPiazza2013}, under the identification
$I_{\rm amp}=I_0$ and the corresponding polarization choice.

\section{Derivation of the dichroic polarization rotation}
\label{app:rotation_derivation}

Let $\delta\equiv\alpha-\theta_g$. In the bright--dark basis of
Eq.~\eqref{eq:bright_dark_modes}, the incident polarization and the
polarization of the transmitted probe are
\begin{align}
	\boldsymbol\epsilon_{\rm in}
	&=-\sin\delta\,\hat{\mathbf e}_b
	+\cos\delta\,\hat{\mathbf e}_d,
	\nonumber\\
	\boldsymbol\epsilon_{\rm out}
	&\propto
	-\sqrt{1-P_b}\sin\delta\,\hat{\mathbf e}_b
	+\cos\delta\,\hat{\mathbf e}_d.
	\label{eq:appB_in_out}
\end{align}
Only the bright component is attenuated because only this component
converts into a real ALP at this order. Here $P_b$ is the conversion
probability for a purely bright input; using the angle-dependent
$P_{\gamma\to\phi}(\alpha)$ in the attenuation factor would count the
bright projection twice.

Writing
$\delta_{\rm out}=\alpha_{\rm out}-\theta_g$ and taking the ratio of
the bright and dark components gives
\begin{equation}
	\tan\delta_{\rm out}
	=\sqrt{1-P_b}\tan\delta,
	\qquad
	\Delta\alpha=\delta_{\rm out}-\delta,
	\label{eq:appB_tangent_exact}
\end{equation}
where the branch of $\delta_{\rm out}$ is chosen continuously such
that $\delta_{\rm out}\to\delta$ as $P_b\to0$.

For $P_b\ll1$, $\sqrt{1-P_b}\simeq1-P_b/2$. Expanding the tangent to
first order gives
\begin{align}
	\tan(\delta+\Delta\alpha)
	&\simeq\tan\delta+\Delta\alpha\sec^2\delta,
	\nonumber\\
	\sqrt{1-P_b}\tan\delta
	&\simeq\tan\delta-\frac{P_b}{2}\tan\delta.
	\label{eq:appB_expansion}
\end{align}
Equating the two sides of Eq.~\eqref{eq:appB_tangent_exact} yields
\begin{equation}
	\Delta\alpha
	=-\frac{P_b}{2}\sin\delta\cos\delta
	=-\frac{P_b}{4}\sin[2(\alpha-\theta_g)].
	\label{eq:appB_rotation}
\end{equation}
The factor $1/4$ arises from two factors of $1/2$: the probability
loss $P_b$ reduces the bright field amplitude by $P_b/2$, while
$\sin\delta\cos\delta=\sin(2\delta)/2$.

Finally, holding the nonpolarization parameters fixed so that $P_b$
is independent of $\alpha$, differentiation of the conversion
probability gives
\begin{equation}
	\frac{\partial P_{\gamma\to\phi}}{\partial\alpha}
	=P_b\sin[2(\alpha-\theta_g)].
	\label{eq:appB_probability_derivative}
\end{equation}
The dichroic rotation is therefore related directly to the
polarization-angle dependence of the conversion probability:
\begin{equation}
	\Delta\alpha
	=-\frac14
	\frac{\partial P_{\gamma\to\phi}}{\partial\alpha}.
	\label{eq:appB_rotation_probability_relation}
\end{equation}
Here $\alpha$ is understood in radians when taking the derivative.
The bright and dark directions have zero rotation, whereas the
rotation magnitude is maximal for
$\alpha-\theta_g=45^\circ$ modulo $90^\circ$. At these angles,
$P_{\gamma\to\phi}=P_b/2$, so maximal rotation occurs when the
incident bright and dark amplitudes have equal magnitude and the
conversion probability is one half of its bright-mode maximum.

\bibliographystyle{apsrev4-2}
\bibliography{references}

\end{document}